\documentclass[reprint,amsmath,amssymb,aps,superscriptaddress]{revtex4-2}

\usepackage{graphicx}
\usepackage{dcolumn}
\usepackage{amsmath}
\usepackage{bm}
\usepackage{mathrsfs}
\usepackage{txfonts}
\usepackage{epstopdf}
\usepackage{subfigure}
\usepackage{caption}
\usepackage{float}
\begin{document}
	
\title{Valley-dependent transport in strain engineering graphene heterojunctions}
	
\author{Fei Wan}
\affiliation{Department of Physics, Hangzhou Dianzi University, Hangzhou, Zhejiang 310018, China}
\author{Xinru Wang}
\affiliation{Department of Physics, Hangzhou Dianzi University, Hangzhou, Zhejiang 310018, China}
\author{Liehong Liao}
\affiliation{Department of Physics, Hangzhou Dianzi University, Hangzhou, Zhejiang 310018, China}
\author{Jiayan Zhang}
\affiliation{Department of Physics, Hangzhou Dianzi University, Hangzhou, Zhejiang 310018, China}
\author{M. N. Chen}
\affiliation{Department of Physics, Hangzhou Dianzi University, Hangzhou, Zhejiang 310018, China}
\author{G. H. Zhou}
\affiliation{Department of Physics and Key
Laboratory for Low-Dimensional Quantum Structures and Manipulation
(Ministry of Education), Hunan Normal University, Changsha 410081,
China}
\author{Z. B. Siu}
\affiliation{Computational Nanoelectronics and Nano-device Laboratory, Electrical and Computer
Engineering Department, National University of Singapore, 4 Engineering Drive 3,
Singapore 117576, Singapore}
\author{Mansoor B. A. Jalil}
\affiliation{Computational Nanoelectronics and Nano-device Laboratory, Electrical and Computer
Engineering Department, National University of Singapore, 4 Engineering Drive 3,
Singapore 117576, Singapore}
\author{Yuan Li}
\email{liyuan@hdu.edu.cn}
\affiliation{Department of Physics, Hangzhou Dianzi University, Hangzhou, Zhejiang 310018, China}
\email[]{Corresponding author: liyuan@hdu.edu.cn}
	
\date{\today}
\begin{abstract}
We study the effect of the strain on the band structure and the valley-dependent transport property of graphene heterojunctions.
It is found that valley-dependent separation of electrons can be achieved by utilizing the strain and on-site energies. In the
presence of the strain, the values of the transmission can be effectively adjusted by changing the strengths of the strain, while the transport angle basically keeps unchanged. When an extra on-site energy is simultaneously applied to the central
scattering region, not only are the electrons of valleys $K$ and $K'$ separated into two distinct transmission lobes in opposite transverse directions, but the transport angles of two valleys can be significantly changed. Therefore, one can realize an effective modulation of valley-dependent transport by changing the strength and stretch angle of the strain and on-site energies, which can be exploited for graphene-based valleytronics devices.
\end{abstract}

\maketitle

\section{Introduction}
Ever since graphene has been experimentally fabricated ~\cite{Novoselov,Geim2007}, extensive attention~\cite{Bolotin2008, Neto2009} has been given to graphene systems owing to the unique band structures and properties of graphene. Pristine graphene is a zero-gap semiconductor and has a linear dispersion relationship near the Dirac points~\cite{Neto2009,Wallace1947}, which causes electrons to behave as relativistic Dirac particles. It has also been experimentally verified that graphene has a remarkably high electron mobility at room temperature, which makes graphene an excellent semiconductor. Because of its exceptional electrical and thermal transport properties, graphene is an important two-dimensional material for the exploration of  physical phenomena in condensed matter ~\cite{Neto2009}, and is expected to be very useful in the next generation of electronic devices.

Many research works have shown that graphene is amenable to external influence by mechanical deformation~\cite{Seon-Myeong,Pereira2009}. Its band structure does not change substantially for realistic strains of up to $15\%$~\cite{Pereira2009,Lee2008}. The effect of long-range strain on electronic properties is a unique feature of graphene~\cite{Suzuura2002, Manes2007}. At low energies, the effects of strains can be modelled as a gauge field that acts as additional pseudo-magnetic field in the momentum operators~\cite{Vozmediano2010}. The most obvious evidence for the unusual effects of strain on the electronic states comes from scanning tunneling microscope measurements of the electronic local density of states of graphene grown on platinum~\cite{Levy2010}. An average compression of $10\%$ creates effective fields of the same order of magnitude as the pseudo-magnetic fields observed in experiments~\cite{Abedpour2011}.
\begin{figure}[b]
	\centering
	\includegraphics[scale=0.6,trim=7 0 0 0, clip]{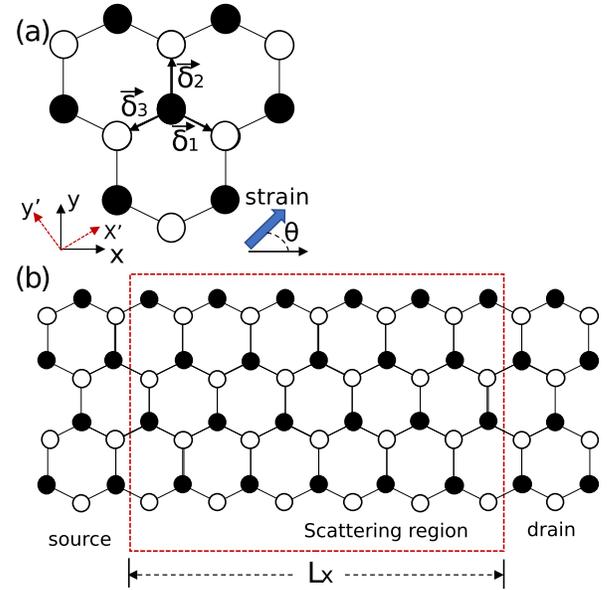}
	
	\caption{(a) Honeycomb lattice geometry. The vectors $\vec{\delta}_{1}$, $\vec{\delta}_{2}$, $\vec{\delta}_{3}$ connect the $A$ sites (white) to their $B$-site (black) neighbors. (b) Schematic of the graphene heterojunction with an applied uniaxial strain in the central scattering region. The zigzag direction of the honeycomb lattice (x-y plane) is parallel to the $x$ axis, and the tension is applied at an angle $\theta$ relative to the $x$ axis.}
	\label{fig:structure}
\end{figure}

Although these previous works have reported many interesting results on the properties of graphene systems under the strain, the effect of the strain on the valley-dependent transport and separation in graphene systems has not been extensively discussed. In this paper, we adopt the tight-binding mode-matching method and propose an efficient way to separate the Dirac fermions in different valleys by utilizing strain and the on-site energy in graphene systems. Our results show that the electrons can be dispersed by strain and the on-site energy in a valley-dependent manner. The combination of strain and on-site energy can be used to realize the effective modulation of valley-dependent transport without the need for ferromagnetic materials or magnetic fields. This phenomenon provides an alternative route to effectively modulate the valley polarizations of graphene devices.

The paper is organized as follows: In Sec.\ref{sec:Model}, we introduce the system under consideration comprising a graphene heterojunction under the influence of strain and an on-site potential applied to the central scattering region. We then calculate the strain-modulated hopping parameters based on the Slater-Koster framework. In Sec.\ref{sec:Results}, we analyze the dispersion relations and employ the mode-matching method to investigate the valley-dependent angular transmission. The combined effects of strain and on-site energies on the valley separation are analyzed and discussed. Finally, a summary is given in Sec.\ref{sec:Conclusions}.

\section{MODEL AND ANALYSIS OF STRAIN}\label{sec:Model}
We consider the dynamics of electrons hopping in the honeycomb lattice governed by the nearest-neighbor tight-binding Hamiltonian~\cite{Song2012,Bahamon2011}
\begin{eqnarray}\label{eq:parameter}
	H&=&\sum_{i}\Omega c^\dag_{i} c_{i}- \sum_{\langle
		i,j\rangle} t(\vec{\delta}_{ij})c^\dag_{i}c_{j},
\end{eqnarray}
where $c_{i}(c^\dag_{i})$ is the electron annihilation (creation) operator on the site $i$, the summation $\sum_{\langle i,j\rangle} $ is only over the nearest-neighbor sites, and $\vec{\delta}_{ij}$ is the vector linking the nearest neighbor sites $i$ and $j$. $\vec{\delta}_{ij}$ takes the values of $\pm\vec{\delta}_1$, $\pm\vec{\delta}_2$, or $\pm\vec{\delta}_3$ where $\vec{\delta}_{1,2,3}$ are shown in Fig. \ref{fig:structure}.

The first term in Eq. (\ref{eq:parameter}) is the on-site potential energy term with $\Omega$ being the magnitude of the on-site energy. The second term is the nearest-neighbor hopping with the hopping energy $t(\vec{\delta}_{ij})$, where the dependence of $t$ on $\vec{\delta}_{ij}$, the vector between the two sites connected by $t$, is explicitly stated to reflect the fact that the applied strain breaks the isotropy of the honeycomb lattice and that of the coupling strengths between a given lattice site and its three neighbors. In the central scattering region, the graphene sheet is uniformly stretched (or compressed) along the angle $\theta$ relative to the $x$ axis. Note that we assume there exists no strain outside the central scattering region.

In the considered Cartesian coordinates, the tension $\mathbf{T}$ can be written as $\mathbf{T}=\mathrm{T}(\cos\theta \hat{e}_x+\sin\theta \hat{e}_y)$.
It is convenient to introduce the principal coordinates $Ox'y'$ in which the tension is written as $\mathbf{T}=\mathrm{T}\hat{e}_{x'}$.
The strain $\epsilon'_{ij}$ is related to the components of the compliance tensor through the generalized Hooke's law~\cite{liyuan2018} by
$\epsilon'_{ij}=\mathrm{T}S_{ijxx}$,
where the indices $i, j=x,y$. For the honeycomb lattice, only the five compliance tensor components $S_{xxxx}$, $S_{xxyy}$, $S_{xxzz}$, $S_{zzzz}$, $S_{yzyz}$ are independent ~\cite{Blakslee1970}.  Thus, the Poisson's
transverse ratio is defined as $\nu=-S_{xxxy}/S_{xxxx}$. For simplicity,  we choose the Possion's ratio to be $\nu=0.165$~\cite{Pereira2009} in our numerical calculations.

It should be mentioned that when stress is mechanically induced in graphene by stretching the substrate, the relevant parameter is in fact the tensile strain rather than the tension $T$. For this reason, we treat $\epsilon$ as the tunable parameter. Because the lattice is oriented with respect to the $Oxy$ axes, the stress tensor needs to be rotated to extract information about the  bond deformations. The strain tensor in the lattice coordinate system reads
\begin{equation} \label{eq:tensor}
	\xi=\epsilon\left(
	\begin{array}{cc}
		\cos^2\theta-\nu \sin^2\theta & (1+\nu) \cos\theta \sin\theta\\
		(1+\nu) \cos\theta \sin\theta & \sin^2\theta-\nu \cos^2\theta\\
	\end{array}
	\right).
\end{equation}

When the strain is applied to the graphene system, the lattice deformation will result changes to the vectors $\vec{\delta}_\ell$ ($\ell=1,2,3$). The strain-dependent vectors are given by $\vec{\delta}_{\ell}=(1+\xi)\cdot\vec{\delta}_{\ell}^0$, which thus
modulate the hopping terms. Accordingly, we obtain the following deformed bond lengths
\begin{eqnarray}
	|\vec{\delta}_{1}|&=&(1+\dfrac{3}{4}\xi_{11}-\dfrac{\sqrt{3}}{2}\xi_{12}+\dfrac{1}{4}\xi_{22})a_0,\\
	|\vec{\delta}_{2}|&=&(1+\xi_{22})a_0,\\
	|\vec{\delta}_{3}|&=&(1+\dfrac{3}{4}\xi_{11}+\dfrac{\sqrt{3}}{2}\xi_{12}+\dfrac{1}{4}\xi_{22})a_0,
\end{eqnarray}
where $\xi_{\kappa,\kappa'} (\kappa=1,2)$ are the matrix components of the strain tensor $\xi$ in Eq.~(\ref{eq:tensor}), and $a_0=0.142\ \mathrm{nm}$ is the distance between nearest neighbor lattice sites in unstrained graphene.
\begin{figure}[t]
	\centering
	\includegraphics[scale=0.55,trim=10 0 0 0,clip]{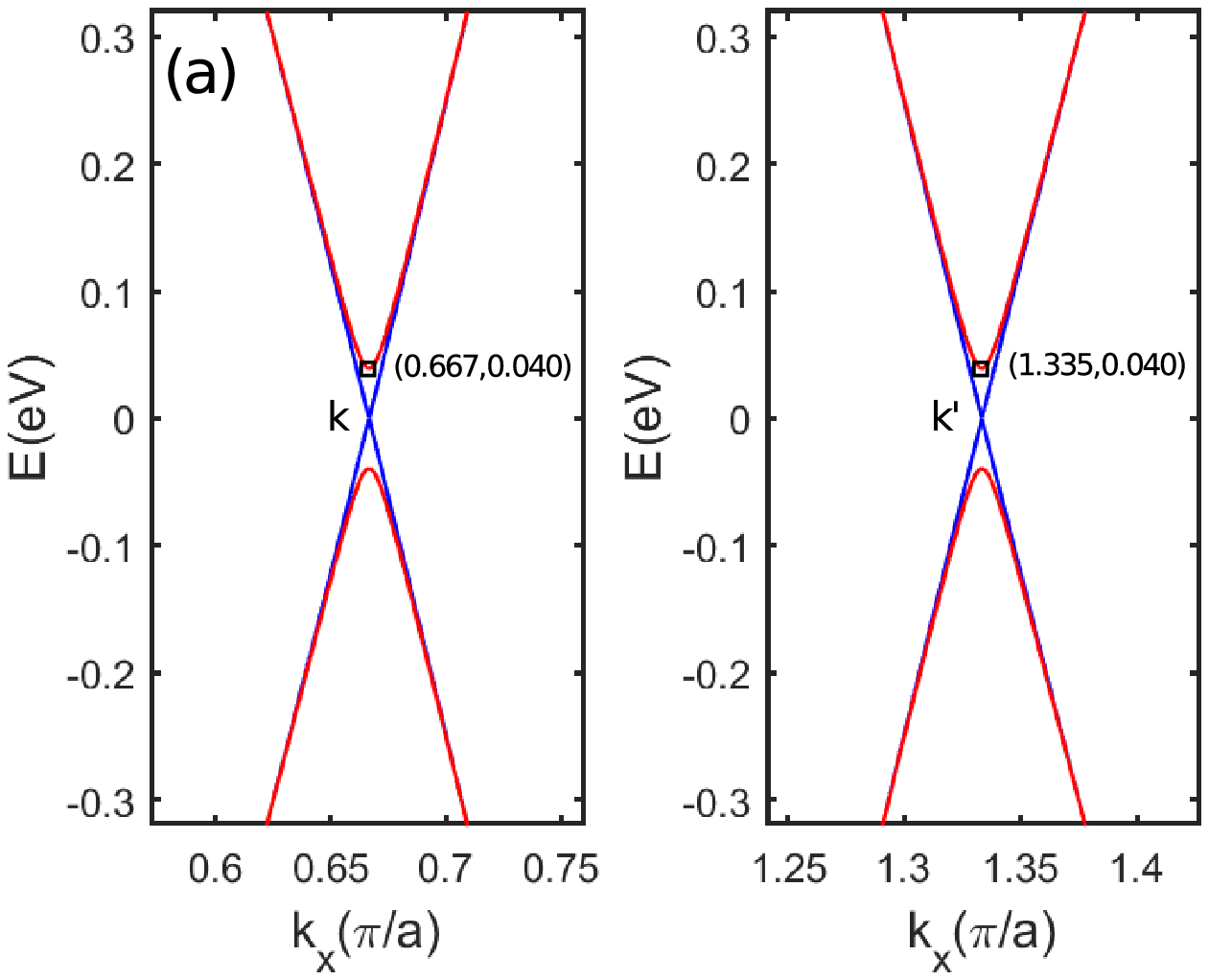}%
	
	\includegraphics[scale=0.55,trim=10 0 0 0,clip]{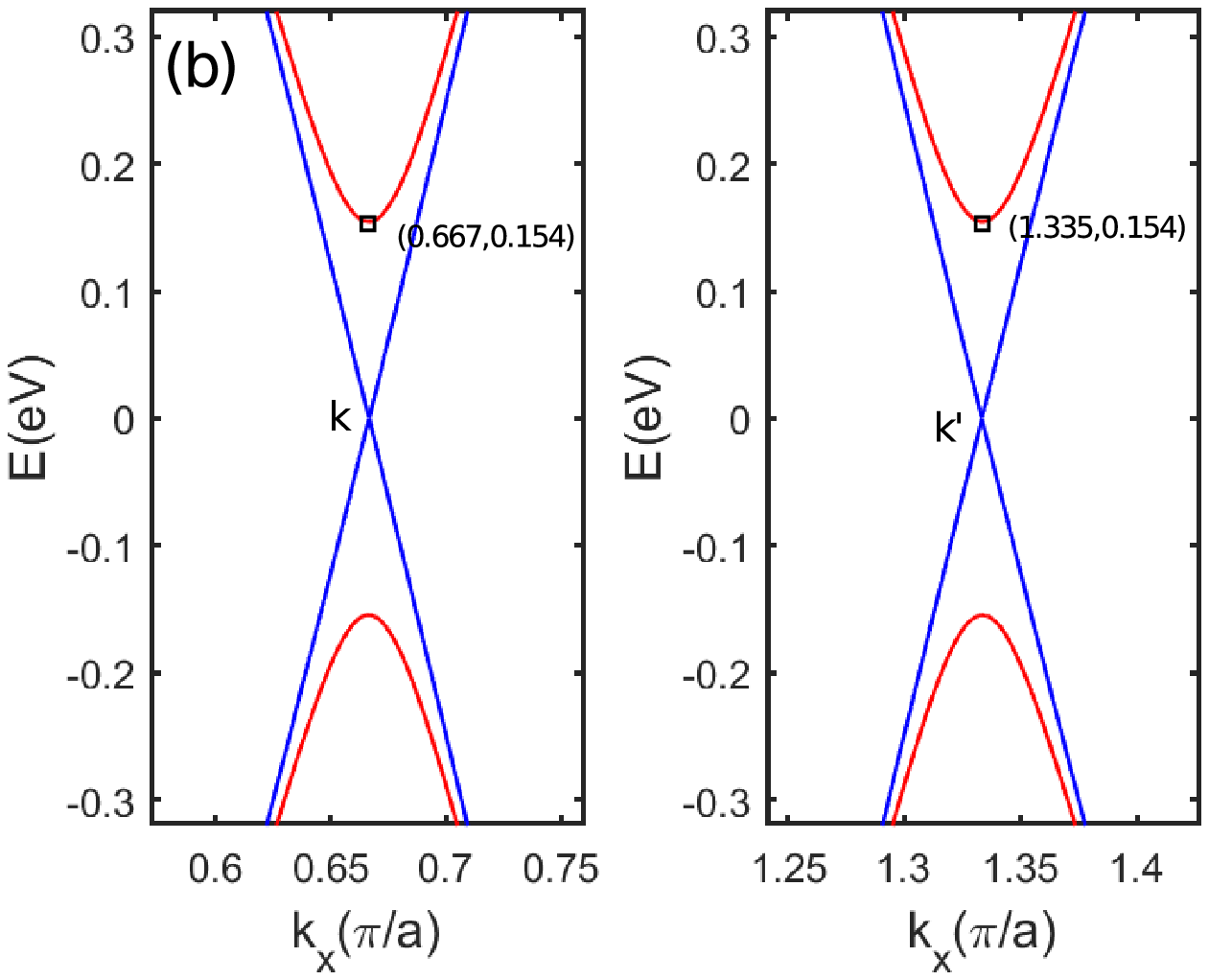}%
	\caption{\label{fig:valley}  Dispersion relations plotted as a function of the wave vector $k_x$ for
		(a) $\epsilon=0.002$, (b) $\epsilon=0.02$.
		The other parameters are $\nu=0.165$, $\theta=45^\circ$, $\Omega=0$ and $k_y=0$. The red (blue) curve
denotes the presence (absence) of the applied strain.}
\end{figure}

The change in bond lengths leads to different hopping amplitudes between each site and its three nearest neighbors.
In the Slater-Koster framework~\cite{Slater1954}, the hoppings can be obtained from the dependence of the integral $V_{pp\pi}$ on the inter-orbital distance. A more convenient assumption is an exponential decay. Thus, we assume that in graphene~\cite{Pereira2009}
\begin{eqnarray}
	t(\vec{\delta_{\ell}})\simeq V_{pp\pi}(r_{\ell})=t_0\exp(-3.37(r_l/a_0-1)),
\end{eqnarray}
where $r_{\ell}=|\vec{\delta}_{\ell}|$  is the bond length, and the hopping energy $t_0$ is $2.7\ \mathrm{eV}$.
We calculate the conductance of graphene using the Landauer formula~\cite{Ando1991,LiYuan2019}
\begin{eqnarray}
	G=\dfrac{e^2}{h}\sum_{m=-N}^NT_m, \hspace{1cm} T_m=\sum_{n=-N}^N|t_{n,m}|^2.
\end{eqnarray}
Here
\begin{eqnarray}
	t_{n,m}=\tilde{\mathbf{\psi}}^\dag_{R,n}(+)G_{S+1,0}[G_{0,0}^{(0)}]^{-1}\psi_{L,m}(+),
\end{eqnarray}
where $\psi_{\mathrm{R}/\mathrm{L},n}(+)$ is the $n$th right-going propagating mode of the right (left) lead, $\tilde{\psi}_{\mathrm{R}/\mathrm{L},n}(+)$ is its dual vector. The source lead is coupled to the system at site 0, and the drain lead at site $S+1$. $G^{(0)}$ and $G$
refer to the Green's functions of the left lead and the full system, respectively, which can be
obtained by using the iterative Green's function approach~\cite{1996Datta,Khomyakov}.
After obtaining the Green's functions, the valley-resolved transmission can be calculated. The $K$ and $K'$ valleys can be distinguished by the  wavevector $k_x$ of the incident source mode $\psi_{L, m}(+)$.   $\psi_{L,m}(+)$ lies in the first valley $K$ if its wave vector is in the range of $k_xa\in (0, \pi)$, and in the second valley $K'$ if its wave vector lies within the range of $k_xa\in (\pi, 2\pi)$~\cite{Cheng2018,Rycerz} where $a=\sqrt{3}a_0$.

Considering periodic boundary conditions along the transverse direction, the transverse crystal momentum $k_y$ can be treated as a parameter in the Hamiltonian, and an infinite graphene sheet can be modeled by using a graphene nanoribbon with the zigzag chain number of $N_y=2$~\cite{liyuan2018,Li2}.
The incident angle is defined as
$\phi=\arcsin(k_y/k_F)$, where the Fermi wave vector $k_F$ can be obtained from the relation $k_F=2E_F/(3a_0t)$.

\section{NUMERICAL RESULTS AND DISCUSSION}\label{sec:Results}
We consider a heterostructure containing a strained central scattering region sandwiched between two unstrained leads in which the interfaces between the leads and the scattering region are along the armchair direction, as shown in Fig.~\ref{fig:structure}. The length of the strained graphene section along the longitudinal direction is denoted as $L_{x}$, where
$L_{x} = [(W_{x}-1)/2] \times\sqrt{3}a_{0}$ and $W_{x}$, the number of distinct rows of atoms along $x$ direction of the sheet, is fixed at $W_{x}=1001$.

\subsection{Influence of strain on band structures}

We investigate the dispersion relation of the infinite-sized homogeneous graphene sheet under the influence of the strain $\epsilon$. The dispersion relations at $k_y=0$ are shown in Fig.~\ref{fig:valley}. In the absence of strain, there is no band gap in the energy band dispersion [see the blue curve]
at $K$  and $K'$ valleys. When $\epsilon=0.002$, the energy difference between the conduction band bottom and valence band top increases to about $80\ \mathrm{meV}$ [see the red curve in Fig.~\ref{fig:valley}(a)].
With the strain strength increases to $\epsilon=0.02$, the energy difference is significantly enlarged to about $308\ \mathrm{meV}$.

\begin{figure*}[t]
	\centering
	\includegraphics[scale=0.5,trim=8 0 20 20, clip]{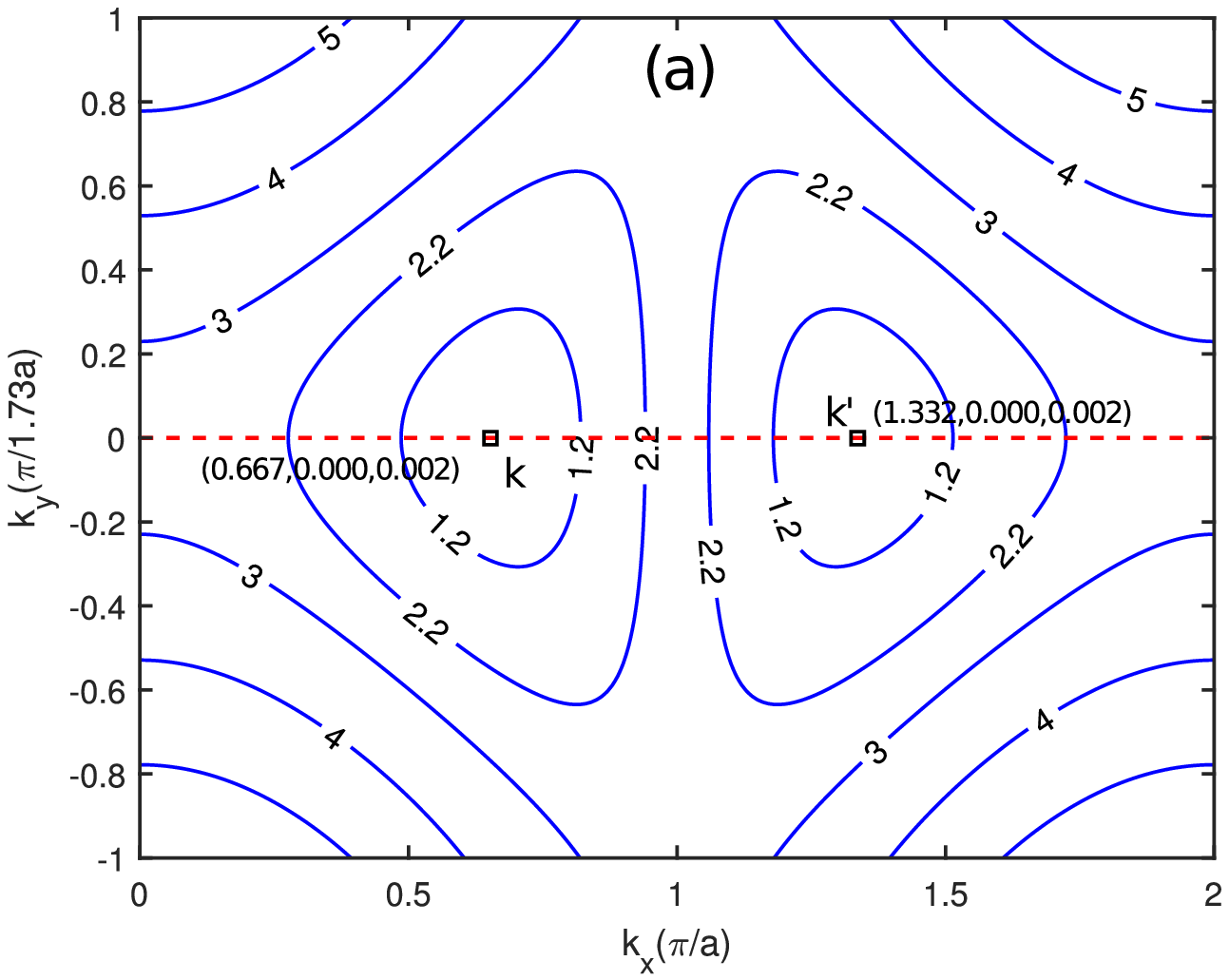}%
	\includegraphics[scale=0.5,trim=8 0 0 20, clip]{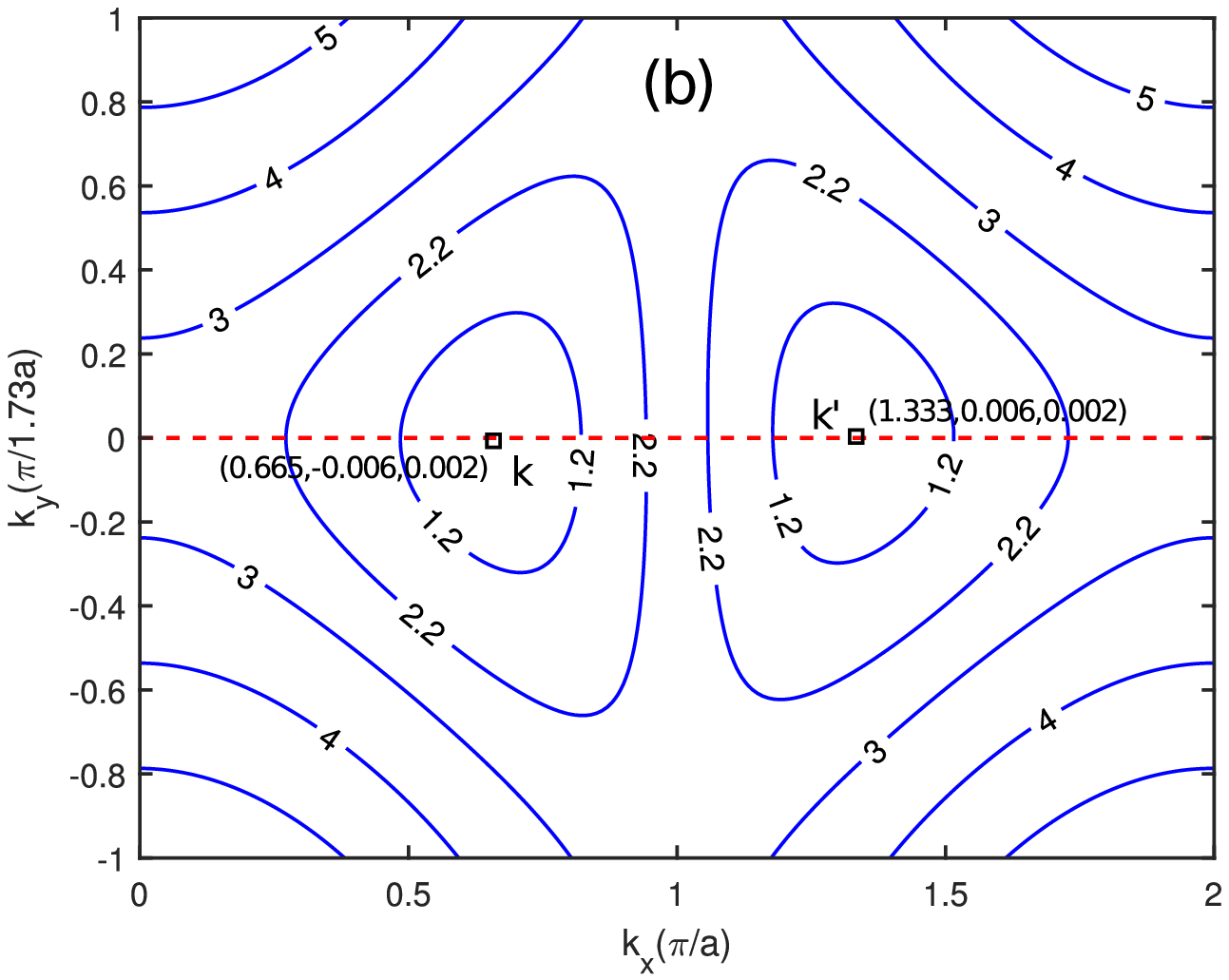}%
	
	\includegraphics[scale=0.5,trim=8 0 20 10, clip]{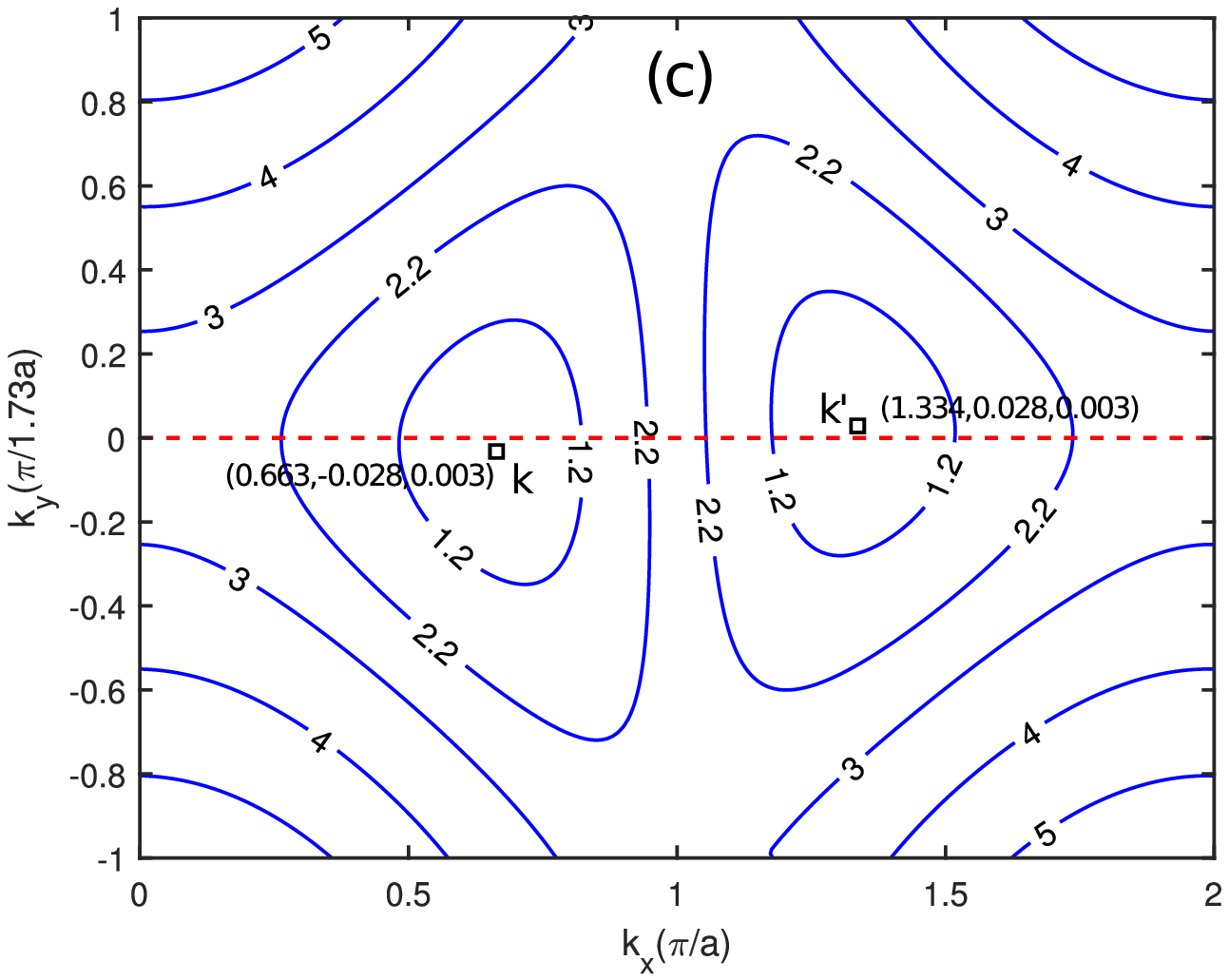}%
	\includegraphics[scale=0.5,trim=8 0 0 10, clip]{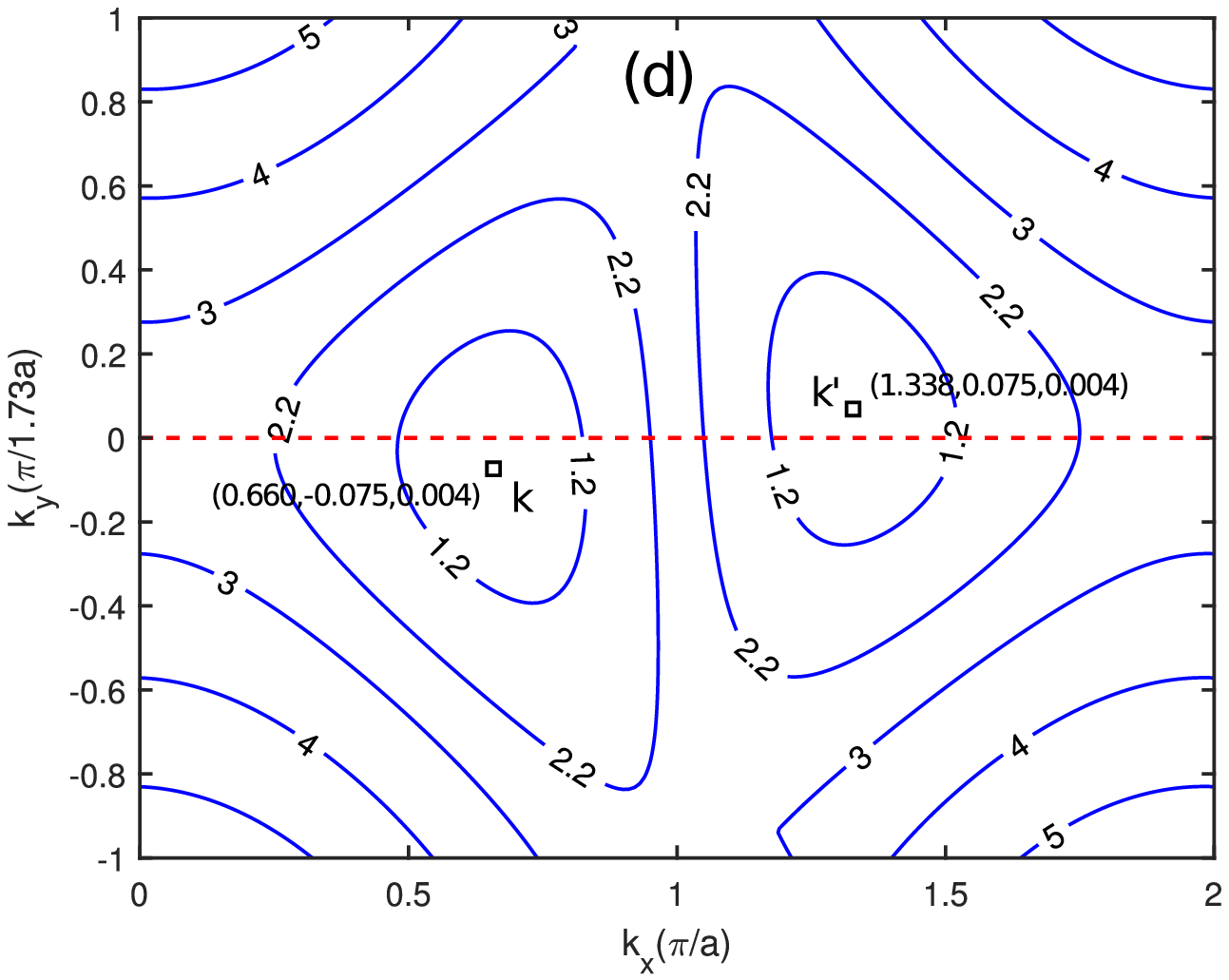}%
	\caption{\label{fig:Coutour} Contour plot of the energies as a function of the wave vectors $k_x$
		and $k_y$ for different strain strengths: (a) $\epsilon=0$, (b) $\epsilon=0.005$, (c) $\epsilon=0.015$, and (d) $\epsilon=0.03$.
		The solid lines are equal energy contour lines corresponding to the energy values of $1.2\ \mathrm{eV}$, $2.2\ \mathrm{eV}$, $3\ \mathrm{eV}$,
		and $4\ \mathrm{eV}$.
		The other parameters are $\Omega=0$, $\nu=0.165$, and $\theta=45^\circ$.
		The black squares denote the $(k_x, k_y, E)$ positions of the Dirac points of the $K$ and $K'$ valleys}
\end{figure*}

It is natural to ask if the minima of the energy profiles for the two valleys still occur at $k_y=0$ in the presence of strain. To clarify the effect of the strain on Dirac points, we plot the equal energy contours as functions of the wave vectors $k_x$ and $k_y$ at different strain strengths in Fig.~\ref{fig:Coutour}. We can see that when $\epsilon=0$, the Dirac points of the $K$ and $K'$ valleys are located at $(k_x,k_y)=(0.667,0)$ and $(1.332,0)$, respectively. When the strain increases from $\epsilon=0$ to $0.005$ in Fig.~\ref{fig:Coutour}(b), the Dirac point of the $K$ valley moves towards smaller values of $k_x$, while the Dirac point $K'$ moves towards the opposite direction . Interestingly, in Fig.~\ref{fig:Coutour}(b)-(d), the Dirac point of the $K$ valley moves towards increasingly negative values of $k_y$, while the Dirac point of the $K'$ valley moves towards larger positive values of $k_y$ as the strain increases from $\epsilon=0$ to $0.005$, $0.015$ and $0.03$. It can thus be seen that the valley separation and the relative transverse shifts of the Dirac cones increase with the strain applied. In particular, the Dirac points of the $K$ and $K'$ valleys also move away from the $k_x$ axis in opposite directions along the $k_y$ direction. Because the dispersion relations in Fig.~\ref{fig:valley}(a) and (b) depict only a cut of the Dirac cone at $k_y=0$, the energy difference between the valence band top and conduction band bottom in the figure does not correspond to the actual band gap.

\subsection{Effect of the strain on the valley-dependent transmission}
We saw in the previous section that the strain can cause the two Dirac points of the strained graphene sheet to move in opposite directions. Here, we first consider the effect of the strain on the transport properties when there is no applied potential in the central scattering section, i.e., when $\Omega=0$ and the graphene sheet is uniformly stretched along the angle $\theta=30^\circ$ relative to the axis $x$.

\begin{figure}[b]
	\centering
	\includegraphics[scale=0.5,trim=100 0 50 0,clip]{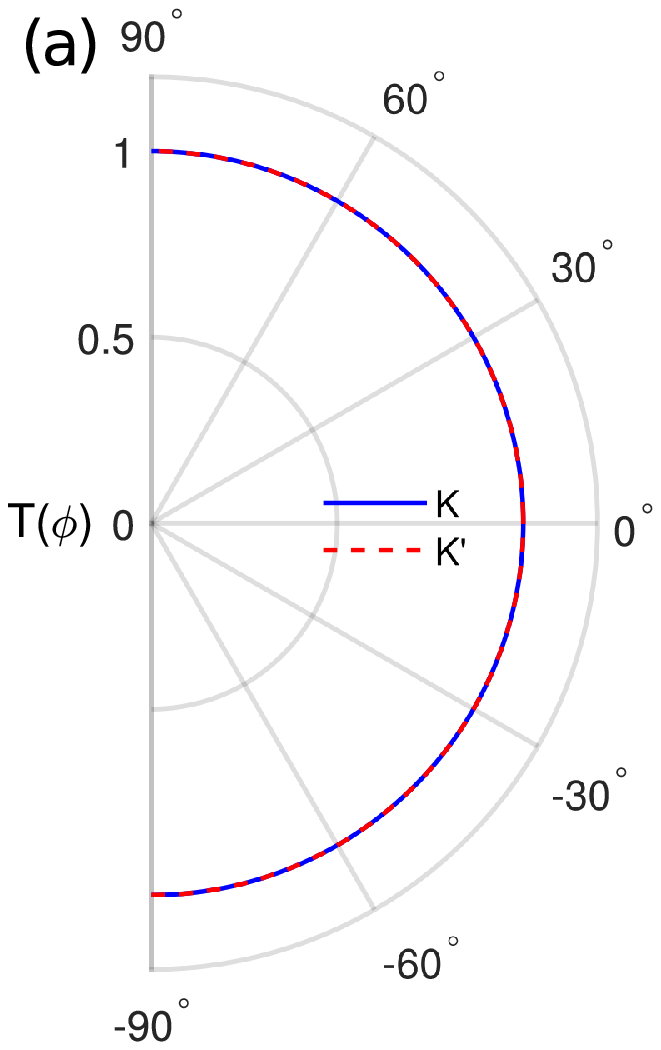}%
	\includegraphics[scale=0.5,trim=100 0 0 0,clip]{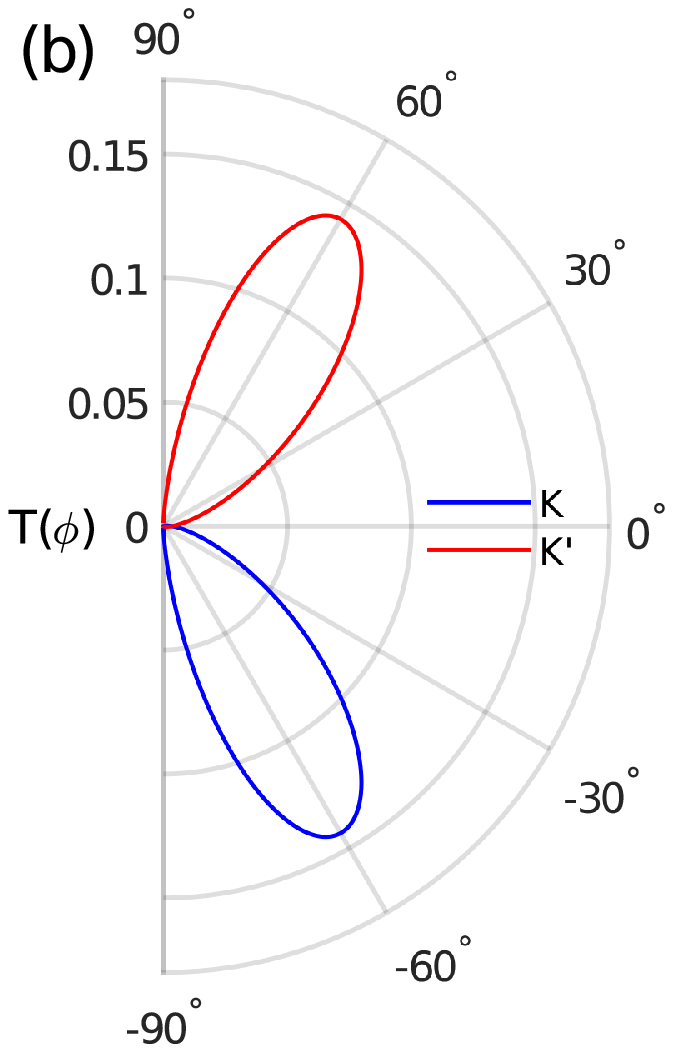}%
	\caption{\label{fig:TransT4} Valley dependence of the transmission plotted as a function of the incident angle $\phi$ for (a) $\epsilon=0$ and (b) $\epsilon=0.003$. The other parameters are $E_F=15\ \mathrm{meV}$, $\theta=30^\circ$, $\Omega=0$ and $L_{x}=123\ \mathrm{nm}$.}
\end{figure}

When $\epsilon=0$ in Fig.~\ref{fig:TransT4}(a), it can be clearly seen that the transmissions of electrons in $K$ and $K'$ valleys are identically equal to 1. This is because the graphene system is homogenous in the absence of any impurities and external disturbances, and the electrons are correspondingly completely transmitted without reflection. When $\epsilon$ is increased to 0.003 in Fig.~\ref{fig:TransT4}(b), the transmission curve of $K$ valley is deflected downwards, while that of the $K'$ valley is deflected upwards. Moreover, the maximum value of the transmission is significantly reduced to about $0.15$ because of the strain. This is because the graphene system is no longer a homogeneous system under the influence of the strain, which leads to the relative shift of two Dirac cones and the non-perfect transmission of electrons in two valleys. These results imply that the strain can be used to separate the Dirac fermions in the $K$ and $K'$ valleys, which is similar to the deflection behavior induced by magnetic fields~\cite{Martino,Masir,Li3}.

\begin{figure}[t]
	\centering
	\includegraphics[scale=0.48,trim=100 0 50 0,clip]{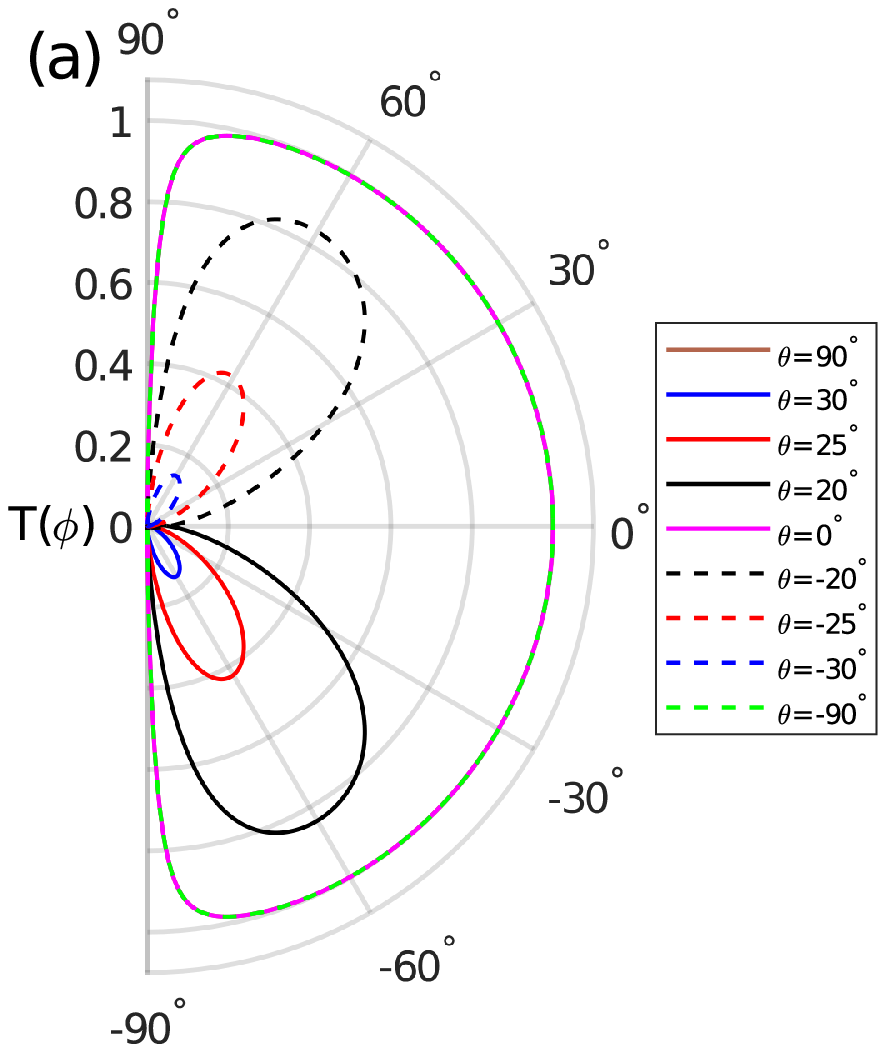}%
	\includegraphics[scale=0.48,trim=100 0 0 0,clip]{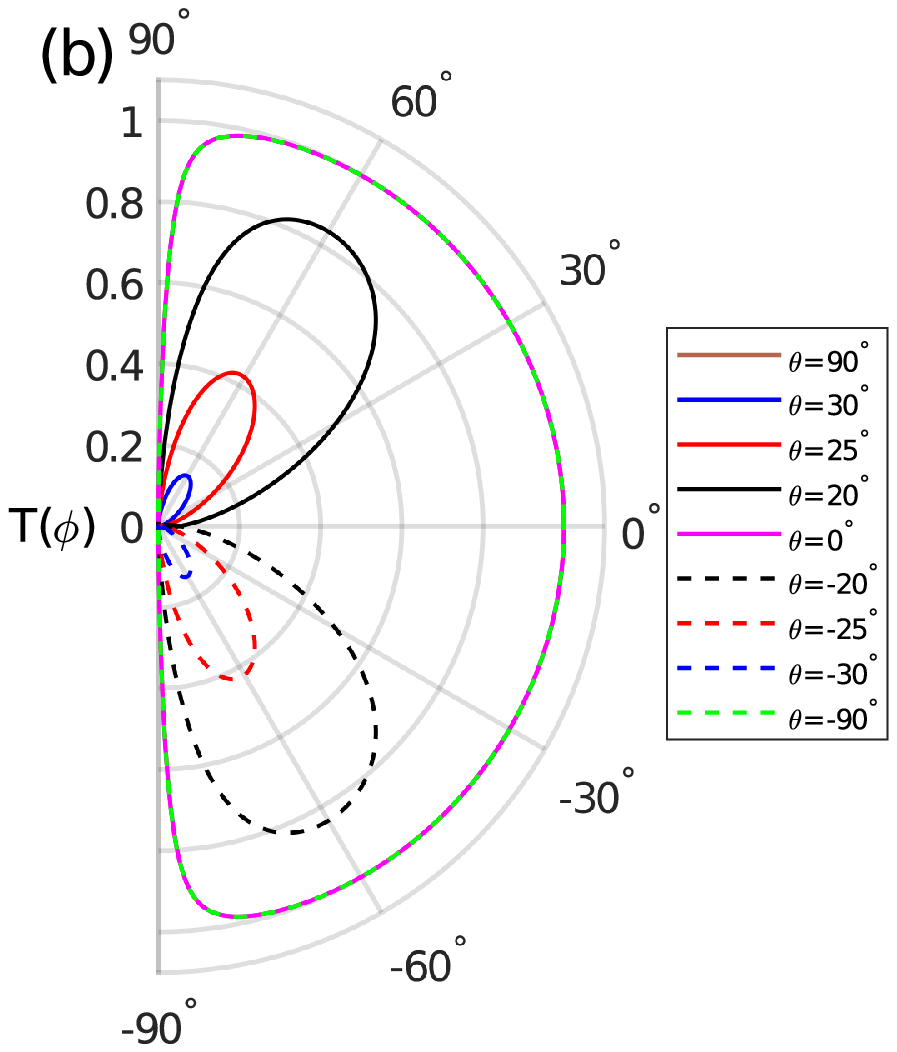}%
\caption{\label{fig:TransT5} The transmission of the (a) $K$ and (b) $K'$ valleys plotted as functions of the incident angle $\phi$ for different strain angles $\theta$. The other parameters are $E_F=15\ \mathrm{meV}$, $\epsilon=0.003$, $\Omega=0$ and $L_{x}=123\ \mathrm{nm}$.}
\end{figure}

In Fig.~\ref{fig:TransT5}, we plot the transmissions of valleys $K$ and $K'$ as functions of the incident angle $\phi$ for different strain angles $\theta$ when $\Omega=0$. It is obvious that the transmission curves of the $K$ ($K'$) valley are deflected downwards (upwards) for positive incident angles from  $\theta=20^\circ$ to $30^\circ$,
as shown in Fig.~\ref{fig:TransT5}. However, as the strain angle increases, the value of the valley-dependent transmission decreases sharply, while the incident angle associating with large values of the transmission basically keeps unchanged. When the angle is changed to negative values of $\theta=-20^\circ$ to $-30^\circ$, the transmission profiles of the $K$ and $K'$ valleys are pushed upwards and downwards, respectively. When the strain is along the zigzag ($\theta=0^\circ$) or armchair direction ($\theta=90^\circ$) $-90^\circ$ , the transmission profiles of two valleys are symmetrical with respective to the normal incident. The electrons in the $K$ and $K'$ valleys therefore cannot be separated at these strain configurations. The results show that the valley-dependent transmission of electrons can be effectively tuned by changing the strain angle in the absence of the on-site energy.

To clarify the effect of the strain strength on the valley-dependent transport when $\Omega=0$, we plot the transmission of the $K$ and $K'$ valleys
as a function of the incident angle $\phi$ for different values of the strain strengths in Fig.~\ref{fig:TransT6}.
We find that the transmission profile of $K$ ($K'$) valley is deflected downwards (upwards) by the strain. As the strain increases from $\epsilon=0.0024$ to $\epsilon=0.003$, the transmission magnitude of the electrons decrease gradually, but the symmetry axes of the transmission lobes are not deflected. The transmission curve of $K$ valley remains within the angle interval $ -75^\circ<\phi<-25^\circ$ [see Fig.~\ref{fig:TransT6}(a)] and that of the $K'$ valley remains within the angle interval $ 25^\circ<\phi<75^\circ$ over the ranges of strain strengths [see Fig.~\ref{fig:TransT6}(b)]. These results imply that when the on-site energy is zero, the transmission profile does not exhibit a deflection behavior, and the valley-dependent transmission of electrons decreases with increasing strain strength.

Based on the analysis above, when $\Omega=0$, strain can effectively cause the separation of the transmission profiles of the electrons in the two valleys, and the valley-dependent transport behavior can be effectively tuned by changing the strength of the strain. However, the symmetry axes of transmission profiles of two valleys remain unchanged when the strain angle or strength of the strain is varied.
\subsection{Combined effect of the strain and on-site energy}
\begin{figure}[b]
	\centering
	\includegraphics[scale=0.45,trim=100 0 50 0, clip]{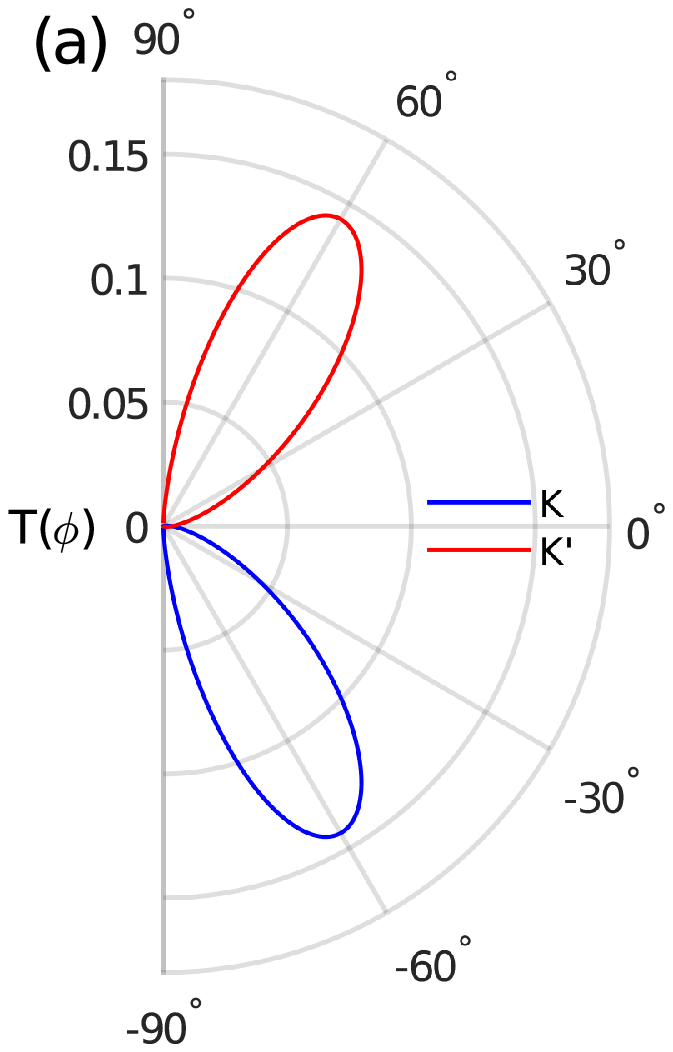}%
	\includegraphics[scale=0.45,trim=100 0 0 0, clip]{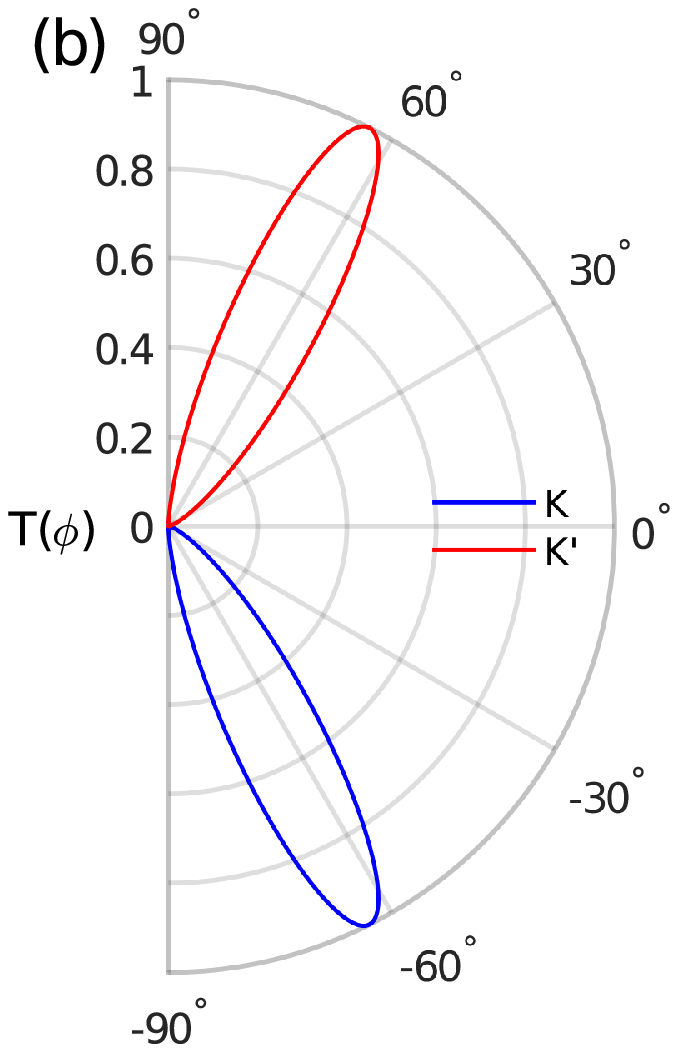}%
	\caption{\label{fig:TransT7} Valley-dependent transmission plotted as a function of the incident angle $\phi$ for (a) $\Omega=0$ and (b) $\Omega=25\ \mathrm{meV}$. The other parameters are $E_F=15\ \mathrm{meV}$, $\theta=30^\circ$, $\epsilon=0.003$, and $L_{x}=123\ \mathrm{nm}$.}
\end{figure}

In the previous section, we saw that the strength of the strain can be used to effectively tune the valley-dependent transmission, but the change in the symmetry axis of the transmission profile is not obvious. We thus further study the combined effect of the strain and the on-site energy on the transport property. In Fig.~\ref{fig:TransT7}(a), when $\Omega=0$, the transmission curve of $K$ ($K'$) valley is deflected downwards (upwards), and the maximum value of the transmission is about $0.15$. In Fig.~\ref{fig:TransT7}(b), when the on-site energy increases to $\Omega=25\ \mathrm{meV}$, the transmission lobe of $K$ valley is centered around $\phi=-63^\circ$, while the transmission lobe of $K'$ valley is centered around $\phi=63^\circ$. Moreover, the addition of the on-site energy significantly increases the maximum value of the transmission to 1. The on-site potential energy results in the formation of a n-p-n type
heterojunction in the central scattering region, which in turn leads to the recovery of perfect Klein tunneling in the considered system. These results imply that the on-site energy can result in the concentration of the electron transport in the $K$ and $K'$ valleys in one special direction and significantly increase the transmission amplitudes, which is similar to the deflection behavior induced by electric fields in silicene systems~\cite{liyuan2018}.
\begin{figure}[t]
	\centering
	\includegraphics[scale=0.48,trim=100 0 50 0,clip]{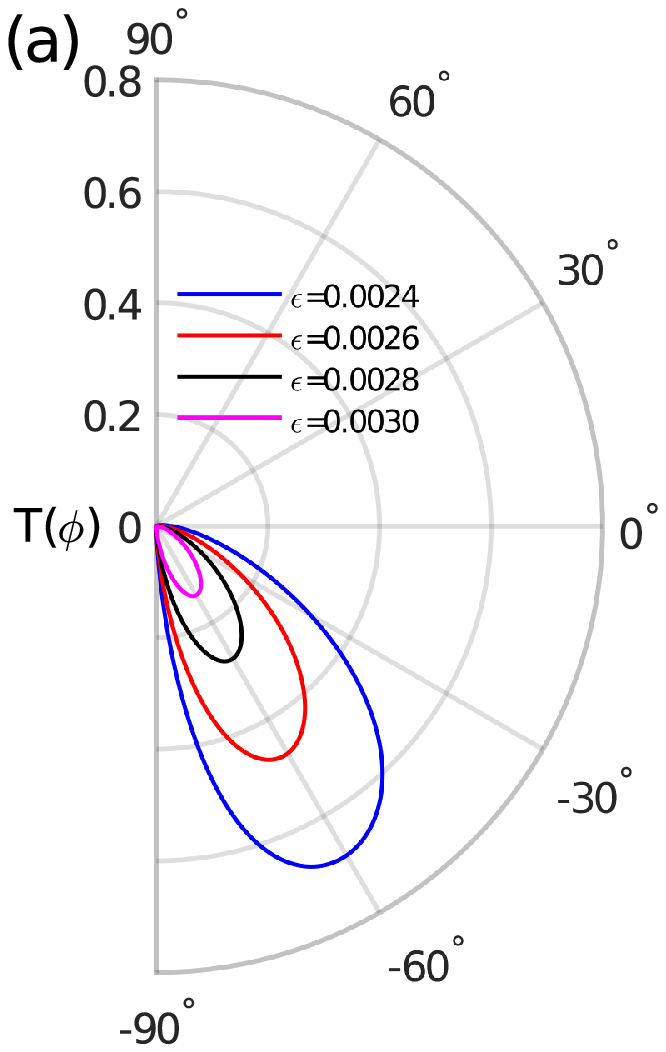}%
	\includegraphics[scale=0.48,trim=100 0 0 0,clip]{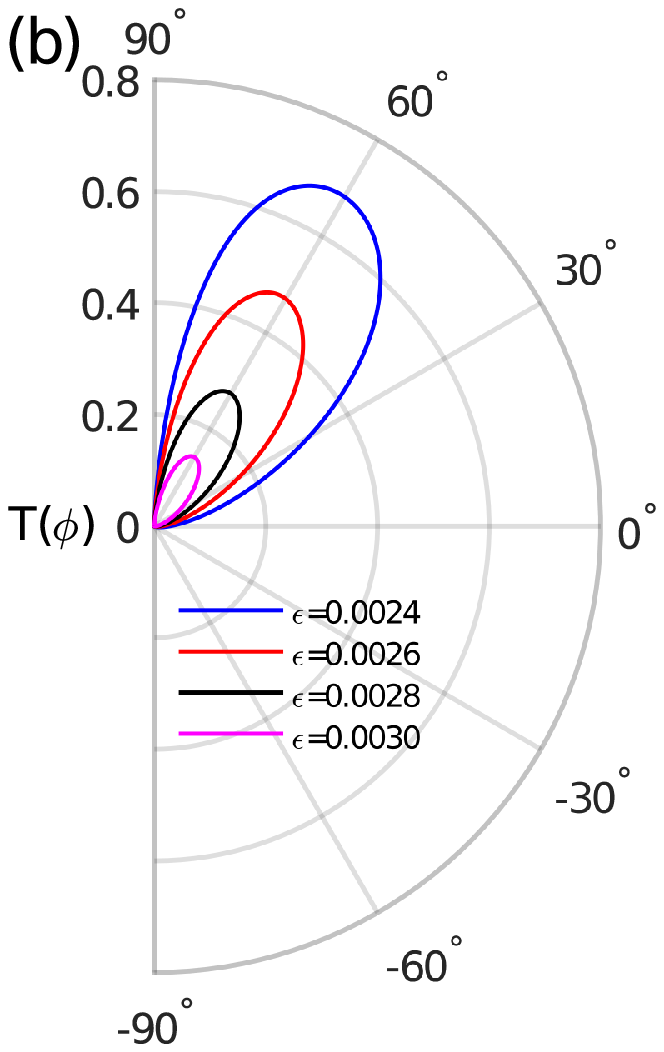}%
	\caption{\label{fig:TransT6} (a) The $K$ and (b) $K'$ valley dependence of the transmission plotted as a function of the incident angle $\phi$ for different strengths of the strain. The other parameters are $E_F=15\ \mathrm{meV}$, $\theta=30^\circ$, $\Omega=0$, and $L_{x}=123\ \mathrm{nm}$.}
\end{figure}
\begin{figure*}[htb]
	\centering
	\includegraphics[scale=0.42,trim=20 0 30 0,clip]{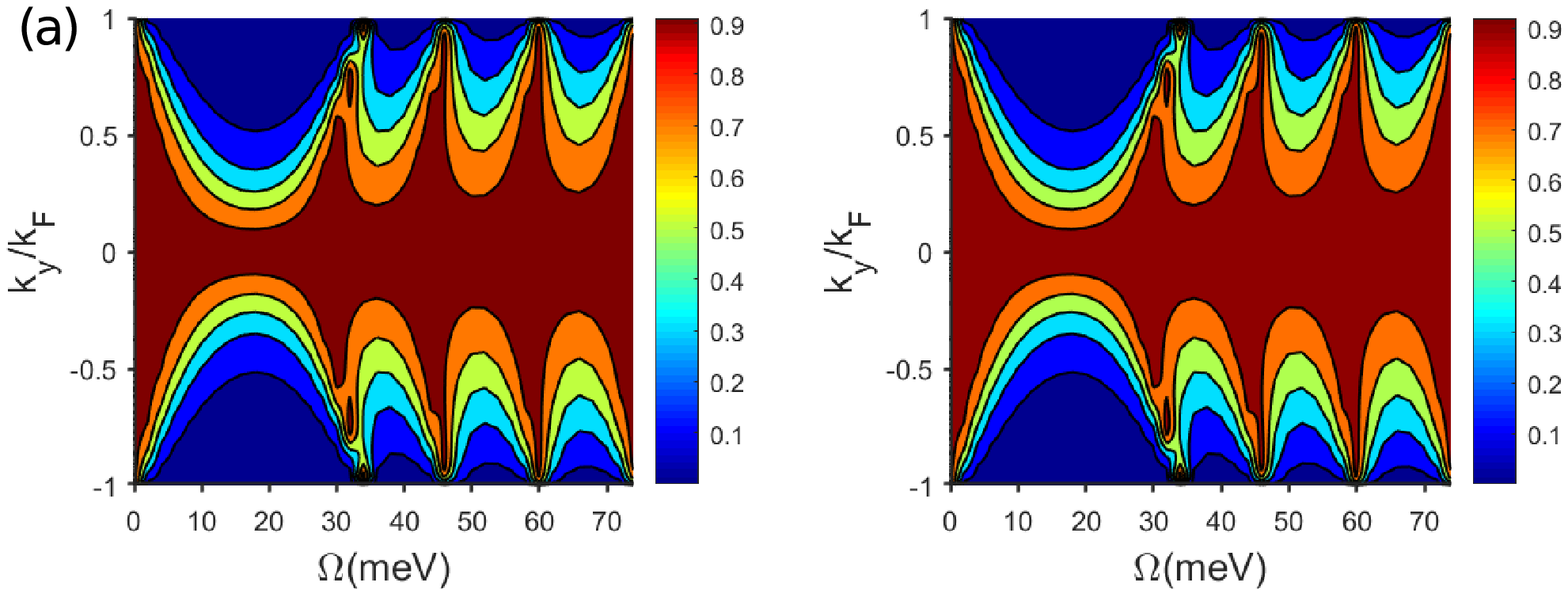}%
	\includegraphics[scale=0.42,trim=20 0 0 0,clip]{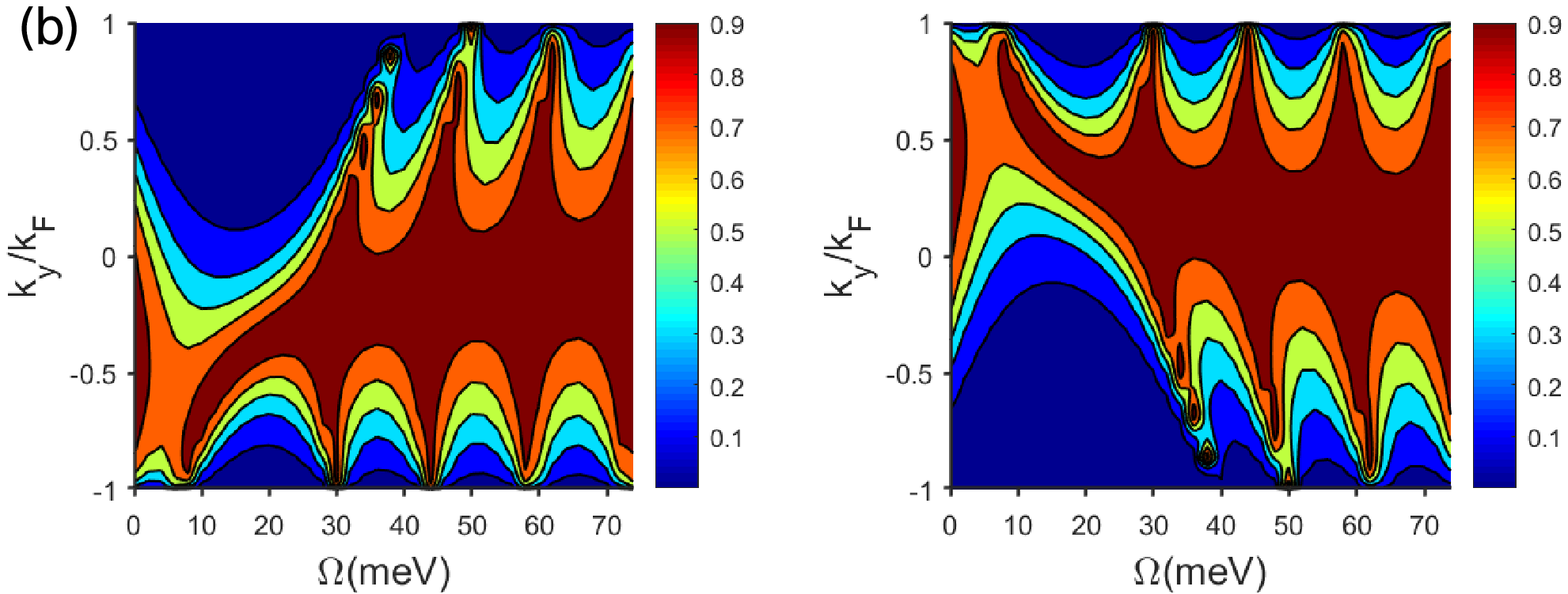}%
	
	\includegraphics[scale=0.42,trim=20 0 30 0,clip]{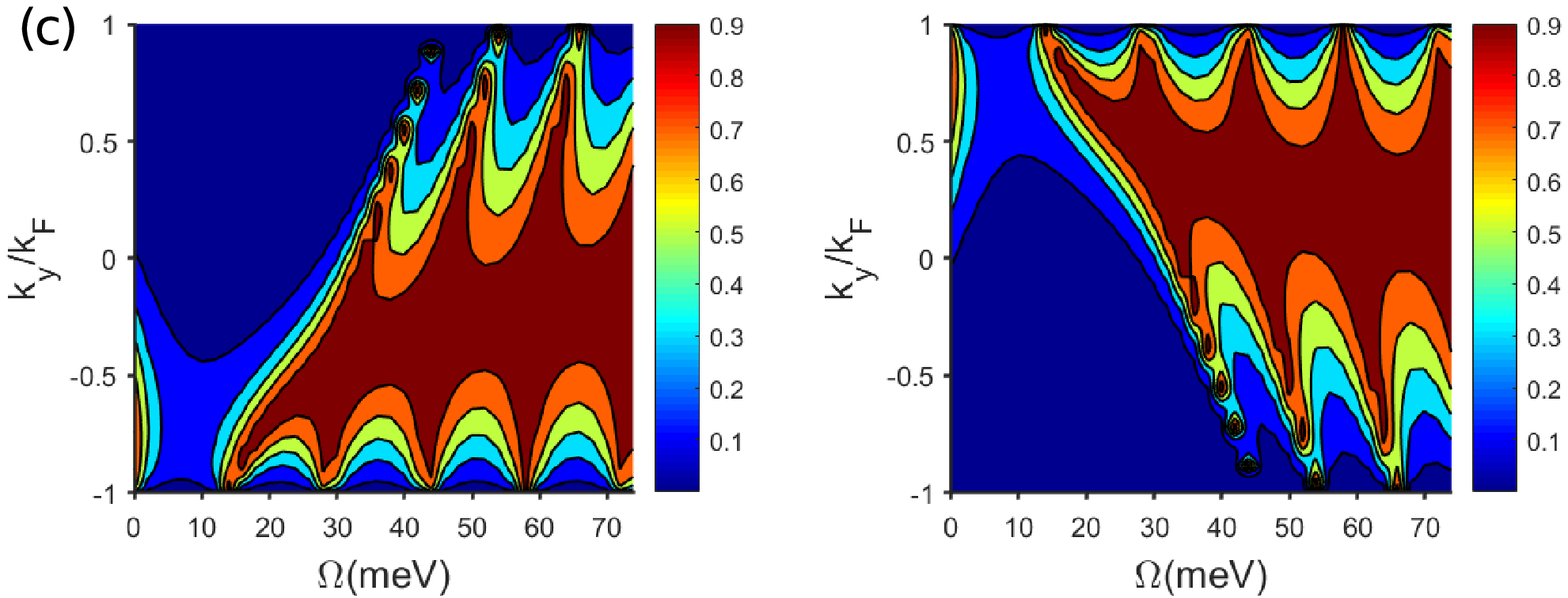}%
	\includegraphics[scale=0.42,trim=20 0 0 0,clip]{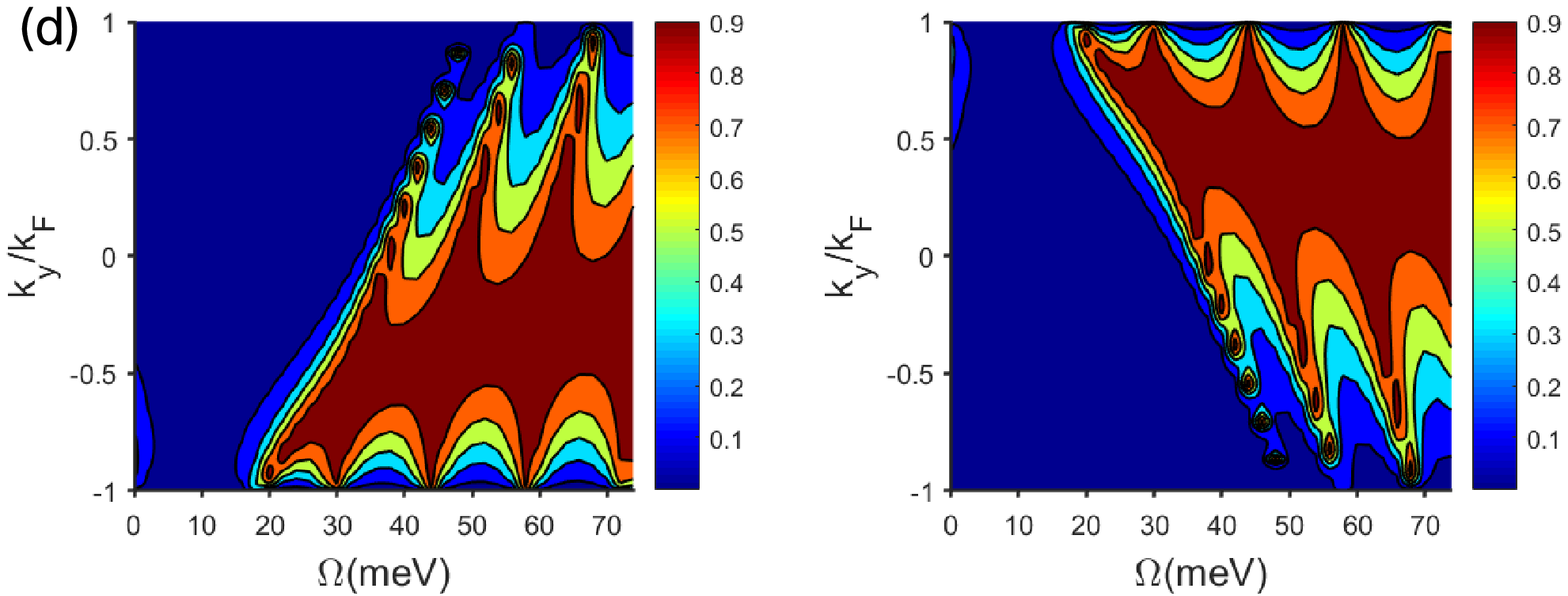}%
\caption{\label{fig:coutour}  Contour plots of the valley-dependent transmission of the graphene heterojunction as a function of the on-site energy $\Omega$ and the wave vector $k_y$ for different strain strengths:
		(a) $\epsilon=0$, (b) $\epsilon=0.001$, (c) $\epsilon=0.002$, (d) $\epsilon=0.003$.
		The other parameters are $\nu=0.165$, $\theta=30^\circ$, and $L_{x}=123\ \mathrm{nm}$. For each panel, the left and right subgraphs are associated with the $K$ and $K'$ valleys, respectively.}
\end{figure*}

To illustrate the effect of the on-site energy on the valley-dependent transport properties,
Fig.~\ref{fig:coutour} shows the contour plots of the transmission of the $K$ and $K'$ valleys as a function of the on-site energy $\Omega$ and the transverse wave vector $k_{y}$ for different strain strengths. For each subgraph in Fig.~\ref{fig:coutour}, the left and right panels are respectively associated with the $K$ and $K'$ valleys. In this numerical calculation, $\theta=30^\circ$ and $E_{F}=15\ \mathrm{meV}$. In Fig.~\ref{fig:coutour}(a), when $\epsilon=0$, electrons in both valleys undergo perfect transmission through the graphene heterojunction at normal incidence irrespective of the values of $\Omega$, which is a characteristic of Klein tunneling. It is obvious that the transmission contour plot is symmetrical about $k_{y}=0$. Interestingly, there exists a region at $10<\Omega<30\ \mathrm{meV}$ in which the incident angle range with significant transmission is contracted. As the strain increases from $\epsilon=0.001$ to $\epsilon=0.003$, the regions with large transmission values for the $K$ valley gradually becomes tilted upwards, while that of $K'$ valley moves downwards. This tilt is particularly obvious in the contraction region within the interval of $10<\Omega<30\ \mathrm{meV}$, which shows a significant effect of the strain and on-site energy.

\begin{figure}[b]
	\centering
	\includegraphics[scale=0.45,trim=100 0 50 0, clip]{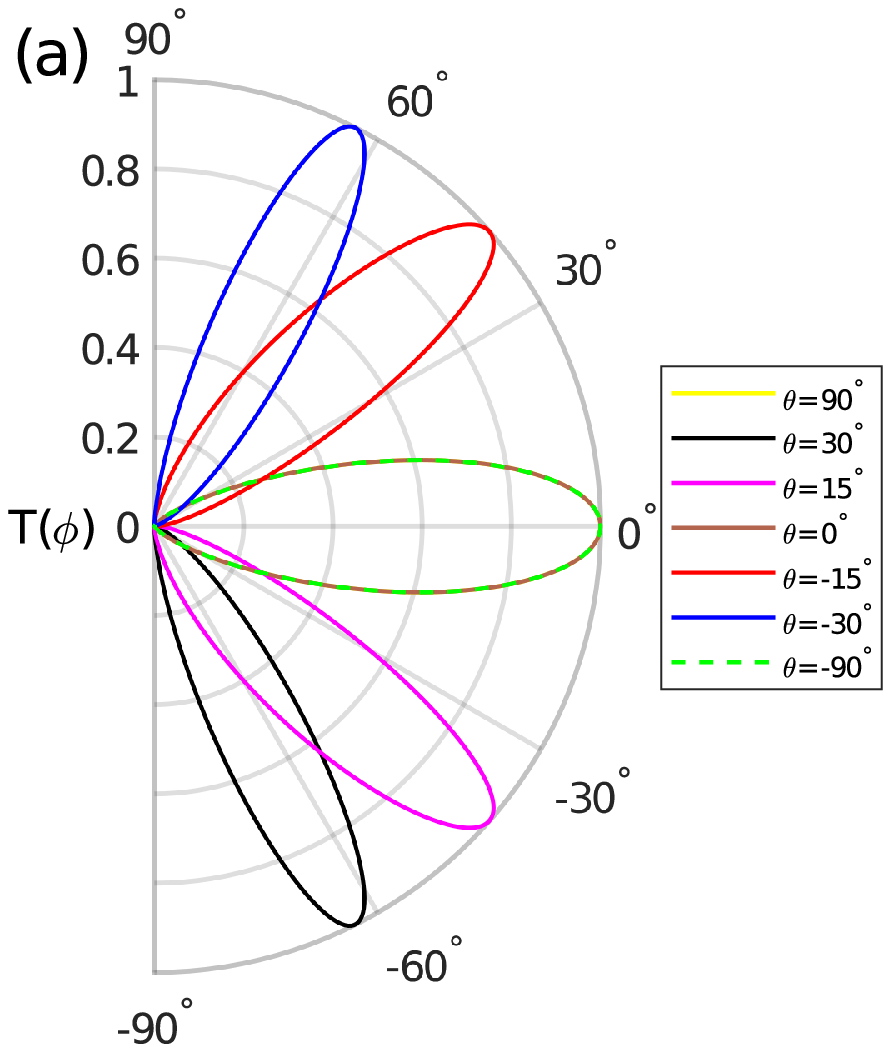}%
	\includegraphics[scale=0.45,trim=100 0 0 0, clip]{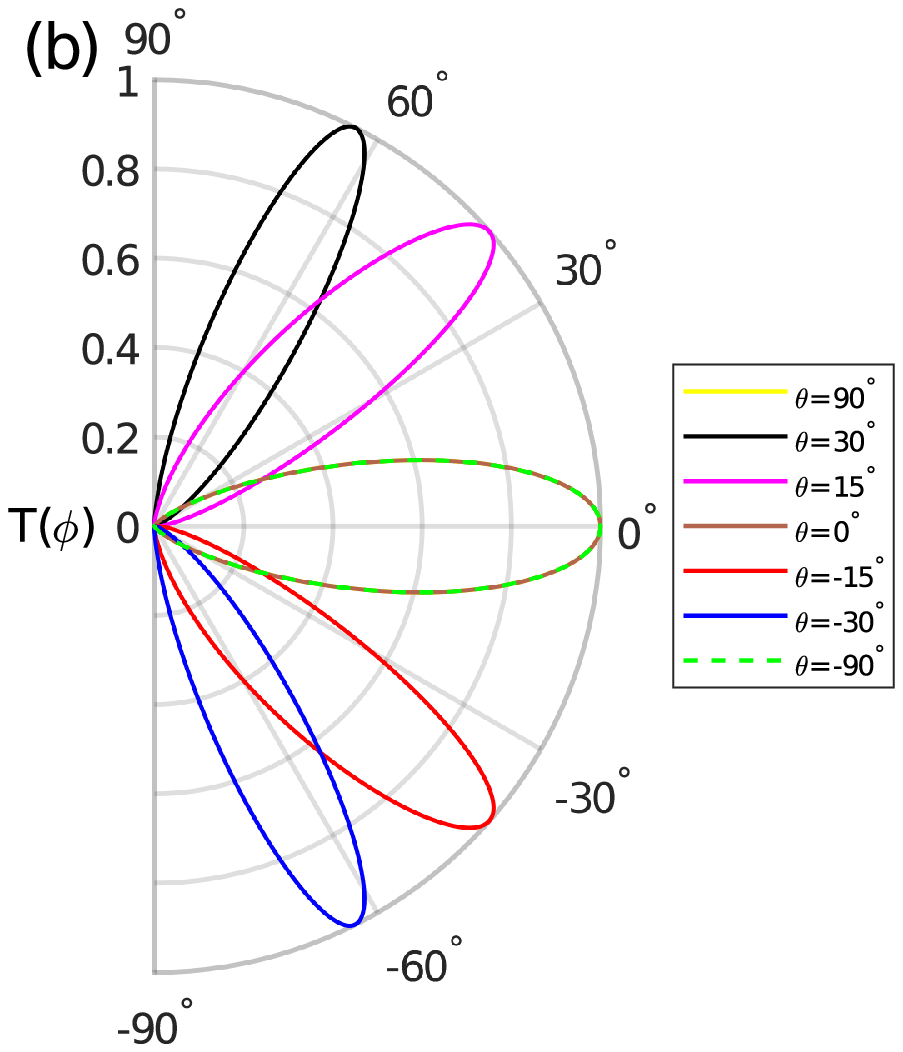}%
	\caption{\label{fig:TransT9} The transmission of the (a) $K$ and (b) $K'$ valleys is plotted as a function of the incident angle $\phi$ for different strain angles $\theta$. The other parameters are $E_F=15\ \mathrm{meV}$, $\epsilon=0.003$, $\Omega=25\ \mathrm{meV}$, and $L_{x}=123\ \mathrm{nm}$.}
\end{figure}

Correspondingly, we further study the effect of the strain angle $\theta$ on the valley-dependent transport at $\Omega=25\ \mathrm{meV}$, as shown in Fig.~\ref{fig:TransT9}. We find that the transmission curves of $K$ ($K'$) valley are deflected downwards (upwards) from $\theta=15^\circ$ to $30^\circ$. However, when the strain angle is changed to negative values, from $\theta=-15^\circ$ to $-30^\circ$, the transmission profiles of $K$ and $K'$ valleys are respectively pushed upwards and downwards, respectively. When the strain angle is increased, the transmission curves shift to larger incident angles,  When the strain is applied along the zigzag or armchair direction, i.e., at $\theta=0^\circ$ or $90^\circ$($-90^\circ$), the transmission profiles of two valleys are symmetrical about $\phi=0^\circ$, and there is no deflection behavior, which is a distinct anisotropy behavior for the strain modulation of the valley current. Thus, the electrons of $K$ and  $K'$ valleys are not separated at these strain configurations. it can be seen that the valley polarization can be effectively modulated by changing the angle of the strain.

\begin{figure}[t]
	\centering
	\includegraphics[scale=0.5,trim=100 0 45 0,clip]{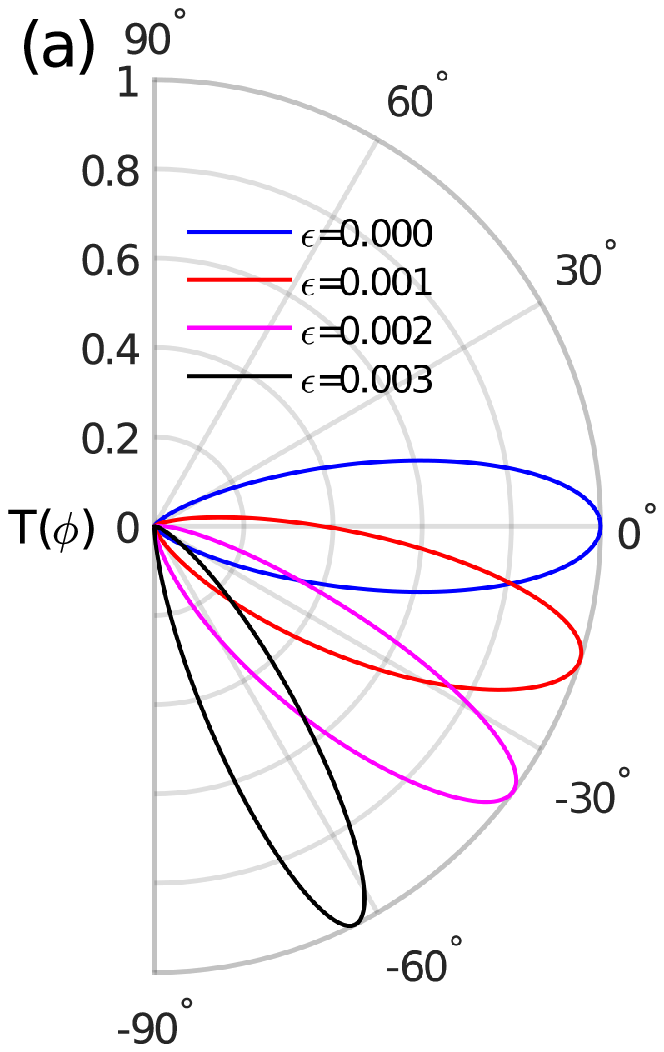}%
	\includegraphics[scale=0.5,trim=100 0 0 0,clip]{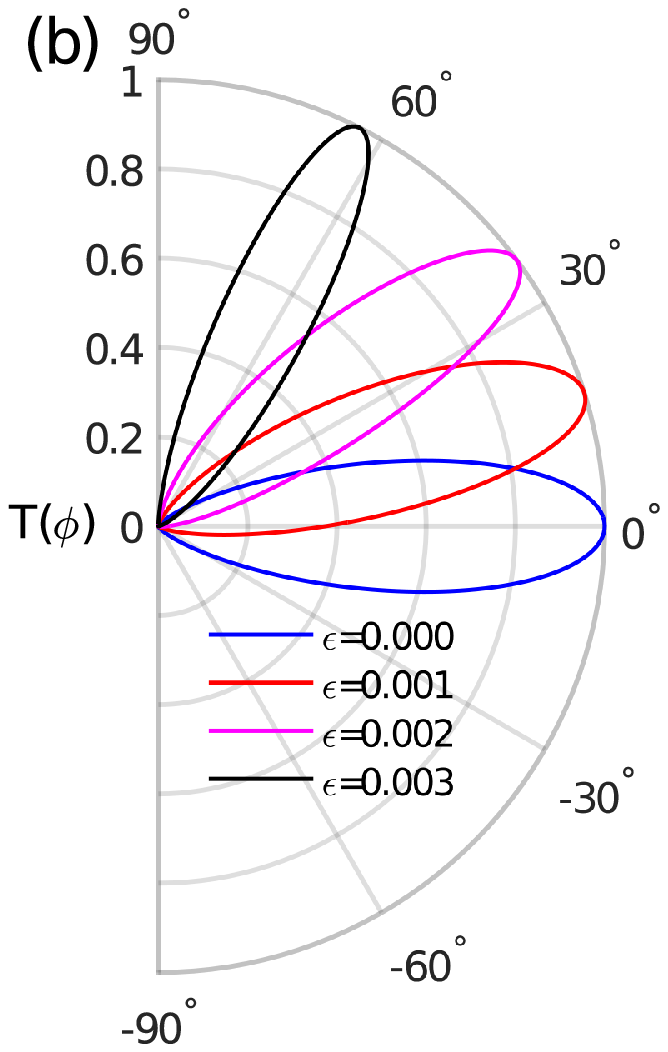}%
	\caption{\label{fig:TransT10} The transmission of the (a) $K$ and (b) $K'$ valleys is plotted as a function of the incident angle $\phi$ for different strengths of the strain. The other parameters are $E_F=15\ \mathrm{meV}$, $\theta=30^\circ$, $\Omega=25\ \mathrm{meV}$, and $L_{x}=123\ \mathrm{nm}$.}
\end{figure}

Similar results can be seen in Fig.~\ref{fig:TransT10}. We plot the transmission of the $K$ and $K'$ valleys as a function of the incident angle $\phi$ for different strengths of the strain when $\Omega=25\ \mathrm{meV}$.
We find that the transmission profile of the $K$ valley is deflected downwards as the strain is increased from $0.001$ to
$0.003$, which differs from the results in Fig.~\ref{fig:TransT6} in which the incidence angle for the peak transmission is independent of the strain magnitude. For example, when $\epsilon=0.002$, the incident angles at which electrons will be transmitted is pushed towards the angular region of $\phi<-30^\circ$ [see Fig.~\ref{fig:TransT10}(a)]. This implies that the $K$-valley electrons will be scattered back into the right region if the incident angle is smaller than a certain critical angle. The transmission profile of the $K'$ valley is correspondingly deflected upwards under the influence of the strain.
The $K'$-valley electrons will be deflected back into the incident region when the incident angle is larger than a certain angle.

Based on the above analysis, when the on-site energy is not zero, strain can be used to effectively separate electrons from different valleys, and the incident angles with high values of the transmission effectively adjusted by changing the angle and strength of the strain. This finding is of great significance for graphene-based valleytronics devices.

\section{CONCLUSIONS}\label{sec:Conclusions}
In summary, we have studied the effects of strain and on-site energy on the dispersion relationship and transport properties of graphene heterojunctions. We found that the electrons can be dispersed in a valley-dependent manner by the uniaxial strain. When the on-site energy is zero and only strain acts on the graphene system, the transmission profiles are deflected to two opposite transverse directions, resulting in the separation of the $K$ and $K'$ valleys. When the strengths of the strain are changed, the magnitude of the transmission is significantly affected by the strain, but the transport angles associated with large values of the transmission are basically keep unchanged. When an additional on-site energy is applied to the scattering region, not only are the electrons in the $K$ and $K'$ valleys separated into two branches, but the transport angle of the electrons are also changed significantly as the angle and strength of the strain increases. Therefore, by combining the strain and on-site energy, an effective modulation of the valley-dependent transport can be achieved by changing the strength and direction of the strain. Our results may be helpful for exploring the transport mechanism of strain-modulated graphene systems and the design of novel types of graphene-based valleytronics devices.

\begin{acknowledgments}
This work was supported by National Natural Science Foundation of China (Grant No. 11574067).
\end{acknowledgments}


\begin{thebibliography}{99}
\bibitem{Novoselov}
K. S. Novoselov,  A. K. Geim, S. V.  Morozov, D. Jiang, Y. Zhang, S. V.Dubonos, I. V. Grigorieva, and A. A. Firsov, \textit{Science} {\bf 306}, 666-669 (2004).
\bibitem{Geim2007}
A. K. Geim and K. S. Novoselov, \textit{Physics Today} {\bf 60}, 35-41 (2007).
\bibitem{Bolotin2008}
K. I. Bolotin, K. J. Sikes, Z. Jiang, G. Fudenberg, J. Hone, P. Kim, and H. L. Stormer, \textit{Solid State Comm.} {\bf 146}, 351-355 (2008).
\bibitem{Neto2009}
A. H. C. Neto, F. Guinea, N. M. R. Peres, K. S. Novoselov, and A. K. Geim, \textit{Rev. Mod. Phys.} {\bf 81}, 109-162 (2009).
\bibitem{Wallace1947}
P. R. Wallace, \textit{Phys. Rev.} {\bf 71}, 622-634 (1947).
\bibitem{Seon-Myeong}
C. Seon-Myeong, J. Seung-Hoon, and S. Young-Woo, \textit{Phys. Rev. B} {\bf 81}, 081407 (2010).
\bibitem{Pereira2009}
V. M. Pereira, A. H. Castro Neto, and N. M. R. Peres, \textit{Phys. Rev. B} {\bf 80}, 045401 (2009).
\bibitem{Lee2008}
C. Lee, X. Wei, J. W. Kysar, and J. Hone, \textit{Science} {\bf 321}, 385-388 (2008).
\bibitem{Suzuura2002}
H. Suzuura and T. Ando, \textit{Phys. Rev. B} {\bf 65}, 235412 (2002).
\bibitem{Manes2007}
J. L. Manes, \textit{Phys. Rev. B} {\bf 76}, 045430 (2007).
\bibitem{Vozmediano2010}
M.A.H. Vozmediano, M.I. Katsnelson and F. Guinea, \textit{Phys. Rep.} {\bf 496}, 109-148 (2010)
\bibitem{Levy2010}
N. Levy, S. A. Burke, K. L. Meaker, M. Panlasigui, A. Zettl, F. Guinea, A. H. Castro Neto, and M. F. Crommie, \textit{Science} {\bf 329}, 544-547 (2010).
\bibitem{Abedpour2011}
N. Abedpour, R. Asgari, and F. Guinea, \textit{Phys. Rev. B} {\bf 84}, 115437 (2011).
\bibitem{Song2012}
J. T. Song, H. W. Liu, H. Jiang, Q. F. Sun, and X. C. Xie, \textit{Phys. Rev. B} {\bf 86}, 085437 (2012).
\bibitem{Bahamon2011}
D. A. Bahamon, A. L. C. Pereira, and P. A. Schulz, \textit{Phys. Rev. B} {\bf 83}, 155436 (2011).
\bibitem{liyuan2018}
Y. Li, Yuan, H. B. Zhu, G. Q. Wang, Y. Z. Peng, J. R. Xu, Z. H. Qian, R. Bai, G. H. Zhou, C. Yesilyurt, Z. B. Siu, and M. B. A. Jalil, \textit{Phys. Rev. B} {\bf 97}, 085427 (2018).
\bibitem{Blakslee1970}
O. L. Blakslee, D. G. Proctor, E. J. Seldin, G. B. Spence, and T. Weng, \textit{J. Appl. Phys.} {\bf 41}, 3373-3382 (1970).
\bibitem{Slater1954}
 J. C. Slater and G. F. Koster, \textit{Phys. Rev.} {\bf 94}, 1498-1524 (1954).
\bibitem{Ando1991}
T. Ando, \textit{Phys. Rev. B} {\bf 44}, 8017-8027 (1991).
\bibitem{LiYuan2019}
Y. Li, W. Q. Jiang, G. Y. Ding, Y. Z. Peng, Z. C. Wen, G. Q. Wang, R. Bai, Z. H. Qian, X. B. Xiao, and G. H. Zhou, \textit{J. Appl. Phys.} {\bf 125}, 244304 (2019).
\bibitem{1996Datta}
S. Datta, Electronic Transport in Mesoscopic Systems (Cambridge University Press, Cambridge, 1995).
\bibitem{Khomyakov}
P. A. Khomyakov, G. Brocks, V. Karpan, M. Zwierzycki, and P. J. Kelly, \textit{Phys. Rev. B} {\bf 72}, 035450 (2005).
\bibitem{Cheng2018}
S. Cheng, H. Liu, H. Jiang, Q. F. Sun, and X. C. Xie, \textit{Phys. Rev. Lett.} {\bf 121}, 156801 (2018).
\bibitem{Rycerz}
A. Rycerz, J. Tworzyd{\L}o, and C. W. J. Beenakker, \textit{Nat. Phys.} {\bf 3}, 172 (2007).
\bibitem{Li2}
Y. Li, Q. Wan, Y. Z. Peng, G. Q. Wang, Z. H. Qian, G. H. Zhou and M. B. A. Jalil, \textit{Sci. Rep.} {\bf 5}, 18458 (2015).
\bibitem{Martino}
A. De Martino, L. Dell'Anna and R. Egger, \textit{Phys. Rev. Lett.} {\bf 98}, 066802 (2007).
\bibitem{Masir}
M. R. Masir, P. Vasilopoulos and F. M. Peeters, \textit{Appl. Phys. Lett.} {\bf 93}, 242103 (2008).
\bibitem{Li3}
Y. Li, M. B. A. Jalil and G. H. Zhou, \textit{Appl. Phys. Lett.} {\bf 105}, 193108 (2014).
\end{thebibliography}
\end{document}